# Modelling Response Time Contrasts in Superconductor Nanowire Single Photon Detectors

Souvik Haldar, Arun Sehrawat, and Krishna B. Balasubramanian

*Abstract*— **Superconducting Nanowire Single Photon Detector (SNSPD) emerges as a potential candidate in the multiple fields requiring sensitive and fast photodetection. While nanowires of low temperature superconducting detectors are mature with commercial solutions, other material options with higher transition temperature and faster responses are currently being explored. Towards this goal, we develop a generalized numerical model that incorporates the thermodynamic properties of the superconducting material and identifies the minimum resolvable photon count for a given bias and device parameters. A phase diagram of detection and latching phases with the minimum number of photons as a function of biasing current and biasing temperature for each material system is presented. We show using the developed model that while low temperature superconducting (LTS) nanowires are more sensitive to the incident photon at different wavelengths, the ultimate limit of a single photon can be achieved using high temperature superconducting (HTS) material such as YBa$_2$Cu$_3$O$_{7-\delta}$, albeit at stringent biasing conditions. On the contrary, ultrafast response time with three orders of magnitude smaller response times can be achieved in select HTS materials making it an appealing for several practical applications.**

## I. INTRODUCTION

Superconducting Nanowire Single Photon Detector (SNSPD) is an important device with applications in multiple disciplines such as light detection and ranging, single-molecule detection, quantum metrology largely due to its near unity system detection efficiencies (SDEs), low dark count rates (DCRs), short timing jitters, high maximum count rates, and photon number resolution capabilities [1,2]. Although numerous research activities were carried out over the past two decades to improve the efficiencies of the SNSPD, choosing the right material systems for designing SNSPD is still an open question. Conventional LTS such as NbN, NbTiN, Nb or WSi are currently employed in many SNSPD applications [2-6] with detection efficiency reaching over 98% at 1550 nm [4-6], dark counts lesser than $10^{-4}$ per second has been observed [7] with time jitters lesser than 100 ps [8]. Nevertheless, ultralow operating temperatures in the range of 120 mK – 5 K requiring liquid He cryogens [3,4-6], and systemic material restrictions leading to longer response times and higher dark count rates has motivated search for other unconventional materials. HTS with short coherence length, intrinsic fast-quasiparticle recombination time, and the higher transition temperatures are recently emerging alternatives for SNSPD applications [2]. However, HTS are also not without issues. Nano structuring the complex composite materials is a challenge. It has been observed that at size scales required for a SNSPD, HTS such as YBa$_2$Cu$_3$O$_{7-\delta}$ experiences inhomogeneity in oxygen stoichiometry [9]. The order parameter degradation, resulting in broadened superconducting transition, reduced critical currents, anisotropy in other physical properties such as thermal conductivity, electron-phonon interactions lead to several complications [10]. In addition, thermally activated phase-slip centers, unbinding of vortex-antivortex pairs, quantum mechanical tunneling of vortices or thermal excitation of single vortices across the HTS nanostripe-edge barrier can also significantly influence the detection mechanism in HTS SNSPD [11]. Thus, with the increasing advent of newer materials and unconventional options for SNSPD devices, there is an increasing need to have a performance prediction for choosing the superconductor for a desired SNSPD response. This requires accurate predictive modelling schemes that can incorporate various aspects of superconducting thermodynamic properties as well as the properties of the incident photon. Such models can also be incorporated in artificial intelligence/machine learning algorithms for material selection.

Superconductors have strong-temperature dependent physical properties that controls the detector response. Different materials tend to show orders of magnitude difference in the SNSPD response times making careful choice crucial. Particularly, in recent experiments HTS show low jitter times and ultra-fast response times in the order of few 100 ps [12-15], while LTS nanowires typically show about 10 – 40 ns. To understand the fundamental operating principles, numerous models, such as normal core hotspot model, diffusion-based hotspot model, photon triggered vortex entry model, diffusion-based vortex entry model, one temperature model, two temperature model etc., have been reported to describe single photon interaction with superconducting nanowires and the emergence of a potential pulse [16]. Of the list, two temperature model (TTM) is widely used to study the hotspot dynamics due to its simplicity and its ability to closely match with experimental results. The

The authors acknowledge the Defense Research and Development Organization (DRDO), Government of India and Industrial Research and Development Unit (IRD), Indian Institute of Technology Delhi, New Delhi 110016, India for financial support.

Souvik Haldar[a], Arun Sehrawat, Krishna B. Balasubramanian are with Department of Material Science and Engineering, Indian Institute of Technology Delhi, New Delhi 110016, India. ([a]email (Corresponding author): ird600280@mse.iitd.ac.in)

models explain several aspects of SNSPD response such as the photon number sensitivity, total response time, timing jitter and the effect of hotspot dynamics on the final response [17]. However, to the best of our knowledge, the crucial influence of superconducting properties on the SNSPD response has been largely ignored. Many of the non-elemental superconductors such as WSi, MgB$_2$, and such, demonstrate radically different dependence of thermodynamic superconducting properties (such a specific heat capacity, thermal conductivity, and inter-particle interaction timescales) with temperature demanding accurate modelling of the particle dynamics to effectively predict device performance. Even for the more mature LTS SNSPDs, a comprehensive model that takes all the thermodynamic properties of the superconductor and that includes the spatio-temporal variations of the photon pulse including the emergence of the normal state resistance and subsequent Joule heating is not yet available. For the emerging HTS materials, no predictive models have been reported yet, thus motivating this research.

In this work we develop a generic theoretical model to illustrate the response contrasts to a single photon from two different superconductors. The model predicts the temperature evolution of the hotspot in a generic superconducting material. Taking the example of Nb for a LTS and YBCO for HTS, we illustrate the effect of material properties and the superconducting parameters on the SNSPD response when the devices is irradiated with a Gaussian photon pulse with two different wavelengths ($\lambda$ = 1550 nm, 830 nm). Using an accurate description of the specific heat capacity, thermal heat capacity, order parameter values and the interfacial particle dynamic rates, we have extracted the variation of electronic temperature, the phonon effective temperatures and finally the overall external potential response for LTS and HTS biased at similar conditions for a single photon pulse. Our calculations match closely with several reported results. We compare and contrast the crucial aspects of photon detection and particle dynamics difference between the two largely different superconducting systems and determine the smallest photon detection capability for a chosen nanowire under different biasing conditions. We observe that while LTS has better photon resolution HTS based SNSPD under similar bias conditions, single photon detection is possible with HTS under slightly different biasing conditions. In contrast, under all biasing conditions, the HTS is at least three orders of magnitude faster than LTS materials. This makes the high temperature unconventional superconductors an important option for future SNSPD devices.

## II. MODEL

A typical detector has a superconducting nanowire realized on a suitable substrate which is biased at an appropriate current density and temperature for maximum sensitivity and low dark count rates. To elucidate the carrier dynamics resulting in a potential signal upon photon incidence, we consider a thermal model accounting for all relevant electron-photon interactions including joule heating. A long meandering nanowire on a substrate where the photon of wavelength, $\lambda$, incidents is shown in the schematic (Fig. 1(A)). The carrier dynamics and the temperature profile of the cross section is solved in a simplified one-dimensional scheme with length ($l$ = 100 μm). The width ($w$ = 50 nm) and thickness ($d$ = 5 nm) is assumed to have no variation in temperature or carrier density as $w, d \ll \lambda$. The schematic diagram of the device geometry as well as the working principle of SNSPD is shown in Fig. 1(A).

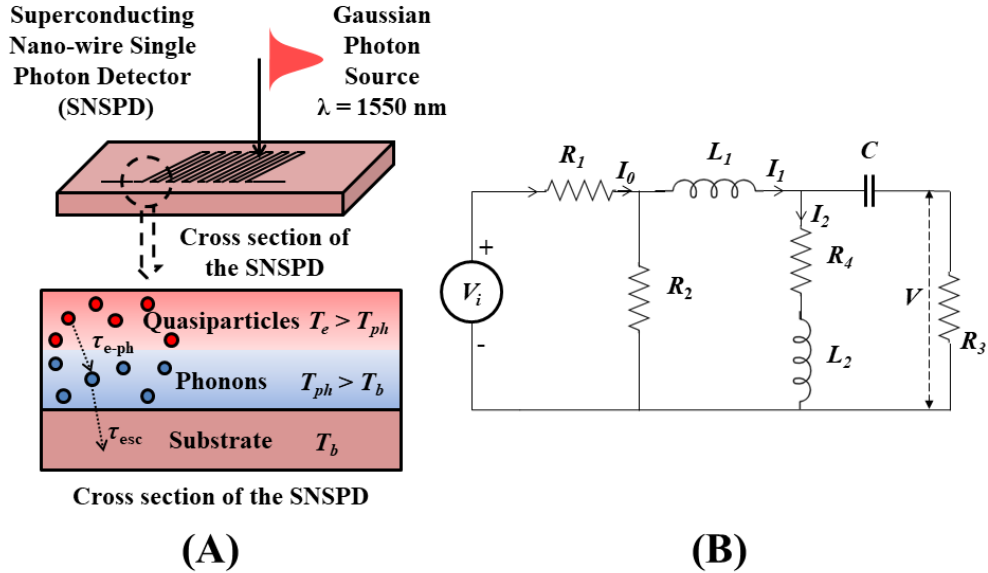

**(A)**                 **(B)**

Fig. 1. (A) Schematic diagram of interaction between a photon pulse and a Superconducting Nano-wire. We have shown a magnified image of a cross section of the nano-wire. The Gaussian photon pulse breaks the Cooper pair and creates energized quasiparticles. The quasiparticles transfer energy to phonon system; eventually the phonons lose the energy through substrate. (B) A practical circuit arrangement for using SNSPD.

The photon of energy greater than the superconducting energy gap initially breaks down the Cooper pairs and the biasing current leads to additional Joule heating due to emergence of resistance in certain parts of the nanowire. The TTM represents the perturbed system using a set of two temperatures, $(T_e)$ and $(T_{ph})$ which are associated with electron and phonon systems respectively [17]. For the electron subsystem, the heat equation can be written as,

$$C_e(T_e)\frac{\partial T_e}{\partial t} - k_e(T_e)\nabla^2 T_e + \frac{C_e(T_e)}{\tau_{e-ph}}(T_e - T_{ph}) = f(x,t)$$

where, $C_e$ is the temperature dependent electronic specific heat, $k_e$ is the electron thermal conductivity, $\tau_{e-ph}$ is the electron-phonon coupling constant. The heat source $f(x,t)$ varies with time and space and accounts for heat generation from incident photon pulse and Joule heating. The electron specific heat $C_e$ of the superconductor varies strongly with the temperature. The thermal conductivity, $k_e$ depends on the specific heat capacity and is given by $k_e(T_e) = D_e C_e(T_e)$, where $D_e$ is the diffusion coefficient of the electron.

The spatio-temporal variance of the heat injected into the nanowire by the photon source $\left(f_p(x,t)\right)$ to be

$$f_p(x,t) = n_p \frac{hc}{2\pi\lambda x_{ph}wd t_{ph}}\exp\left(-\frac{(x-x_0)^2}{2x_{ph}^2}\right)\exp\left(-\frac{(t-t_0)^2}{2t_{ph}^2}\right)$$

where $n_p$ is the number of photons at the designated wavelength $\lambda$, $\frac{hc}{2\pi\lambda x_{ph}wd t_{ph}}$ is the height of the curve peak, $x_0$ is the spatial position of the center of the peak, $x_{ph}$ is the spatial standard deviation or Gaussian root mean square width, $t_0$ is the temporal position of the center of the peak and $t_{ph}$ is the temporal standard deviation.

With the onset of electrical resistance upon photon incidence, the Joule heating is included in the source term as $f_j(t) = J_b^2\rho(T_e)$, where, $\rho(T_e)$ is the temperature dependent resistivity of the superconductor.

Following the electron subsystem, the phonon subsystem equation can be written as below,

$$C_{ph}(T_{ph})\frac{\partial T_{ph}}{\partial t} - k_{ph}(T_{ph})\nabla^2 T_{ph} - \frac{C_e(T_e)}{\tau_{e-ph}}(T_e - T_{ph}) + \frac{C_{ph}(T_{ph})}{\tau_{esc}}(T_{ph} - T_{sub}) = 0$$

Where $C_{ph}$, $k_{ph}$ are phonon specific heat, and phonon thermal conductivity. The heat exchange with the electron sub-system is expressed via the $\tau_{e-ph}$ and the heat loss to the substrate is expressed via the phonon escape time to the substrate with a rate proportional to $\frac{1}{\tau_{esc}}$. The substrate temperature is taken to be $T_{sub}$. The electron and phonon subsystems equations are solved simultaneously to obtain the relevant time and spatial dependence of the $T_e$ and $T_{ph}$.

Using the temperature evolution based on the photon incidence, we obtain the temporal characteristics of the detector electrical signals. The nanowire resistance emergence leads to potential pulses in the shunt resistor according to the electrical circuit in Fig. 1(B). The total kinetic inductance of the superconducting nanowire is extracted from the nanowire geometry using $L_k(T) = \mu_0\lambda^2(T)\left(\frac{l}{A}\right)$, where $\mu_0$ is the magnetic permeability in vacuum, $\lambda(T)$ is the London penetration depth, $A$ is the area of the nanowire geometry [18]. The governing equation for the electrical response is written as below (refer supplementary material **Section 1** for further details).

$$\begin{bmatrix}\frac{dI_1}{dt}\\\frac{dI_2}{dt}\\\frac{dV}{dt}\end{bmatrix} = \begin{bmatrix}-\frac{(R_3+R_2)}{L_1} & \frac{R_3}{C} & -\frac{1}{L_1}\\\frac{R_3}{L_2} & -\frac{(R_3+R_4)}{L_2} & \frac{1}{L_2}\\\frac{1}{C} & -\frac{1}{C} & 0\end{bmatrix}\begin{bmatrix}I_1\\I_2\\V\end{bmatrix} + \begin{bmatrix}\frac{I_0R_2}{L_1}\\0\\0\end{bmatrix}$$

The model developed till now is generic. Custom functions for various superconducting properties such as $K_e, C_e, C_{ph}, K_{ph}, \tau_{esc}$ and $\Delta(T)$ can be defined for specific superconductor permitting response calculations. Custom finite difference numerical codes using Crank-Nicholson scheme was employed to solve the coupled equations and obtain rise, fall and the response time of the nanowire for varying photon number and superconducting material properties.

## III. RESULTS

We first calculate the response from a conventional LTS. Parameters of Nb is taken here as a representative material. The superconducting thermodynamic variables of niobium for simulating the equations **1** and **4** are listed in the following TABLE I.

### TABLE I
#### MACROSCOPIC THERMODYNAMIC SUPERCONDUCTING VARIABLES OF LTS AND HTS

| Thermodynamic variables | Symbols | Values | |
|---|---|---|---|
| | | **Nb** | **YBCO** |
| Transition temperature | $T_c$ | 9.6 K | 92 K |
| Critical current density ($H = 0, T = 0$) | $J_c$ | $6 \times 10^{10}$ A/m² [19] | $1 \times 10^{12}$ A/m² [24] |
| Normal state resistivity | $\rho_N$ | $1 \times 10^{-7}$ Ωm [20] | $1 \times 10^{-6}$ Ωm [25] |
| Superconducting energy gap | $\Delta$ | $1.76 k_B T_c \sqrt{1 - \frac{T}{T_c}}$ J … $T < T_c$ [21] | 16.3 meV at $T = T_c$ [26] |
| Electronic heat capacity | $C_e$ | $0.92 \times 10^6 \exp\left(-\frac{\Delta(T)}{k_B T}\right)$ J/m³K … $T < T_c$ [21] $0.058 \times 10^6 T$ J/m³K … $T > T_c$ [21] | $425.628 \frac{T^3}{T_c^2}$ J/m³K … $T < T_c$ [27] $141.876 T$ J/m³K … $T > T_c$ [27] |
| Phonon heat capacity | $C_{ph}$ | $9.8 T^3$ J/m³K [21] | $3.84 T^3$ J/m³K [28] |
| Electron diffusion coefficient | $D_e$ | $1 \times 10^{-4}$ m²/s [21] | $4 \times 10^{-5}$ m²/s [29] |
| Phonon diffusion coefficient | $D_{ph}$ | $1 \times 10^{-5}$ m²/s [21] | $1.5 \times 10^{-5}$ m²/s [30] |
| Electron – phonon coupling constant | $\tau_{e-ph}$ | $2\left(\frac{6.5}{T}\right)^2$ ns [22] | $1 - 2$ ps at $80 - 90$ K [31] |
| Phonon escape to substrate | $\tau_{esc}$ | 40 ps [22] | $R_b C_{ph}(T) d$ where, $R_b = 5 \times 10^{-8}$ m²K/W is the thermal boundary resistance between YBCO and MgO substrate. [31] |

The prominent discontinuity for the electronic specific heat closes to $T_c$ is noted; while the phonon thermodynamic properties show conventional variation with the temperature, $C_{ph} \propto T^3$. The detailed variations of $C_e(T_e), C_{ph}(T_{ph}), k_e = D_e C_e(T_e)$ and $k_{ph} = D_{ph} C_{ph}(T_{ph})$ are given in **supplementary material Section 2**.

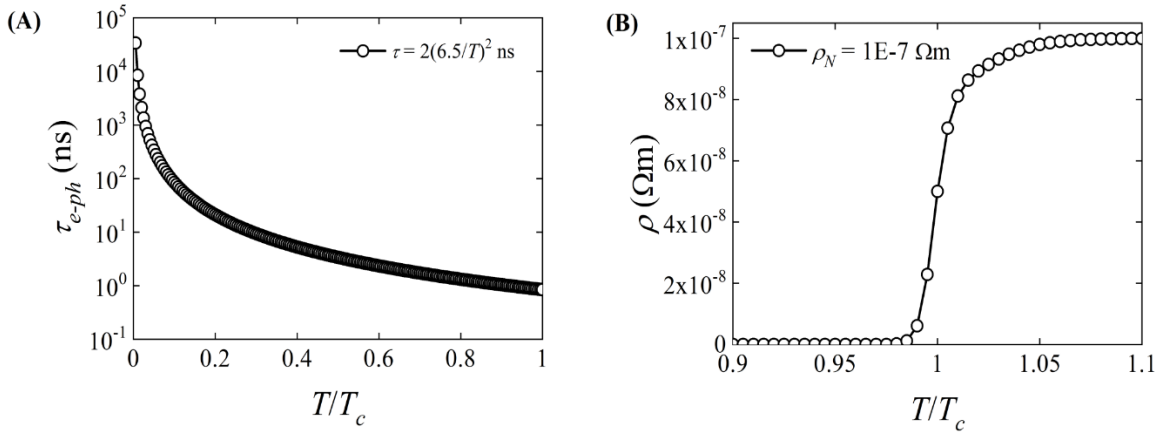

Fig. 2. (A) Variation of electron-phonon coupling constant, $\tau_{e-ph}$ with normalized temperature $T/T_C$, the variation of $\tau_{e-ph}$ has been represented in log scale. (B) Variation of resistivity, $\rho$ of LTS with $T/T_C$; the fit parameters are $a = 0.0583$ K, $b = 0.194$ K and $c = 0.263$ K.

In clean ultrathin (< 20 nm) superconducting films electron-phonon interaction time ($\tau_{e-ph}$) is observed to vary as $\tau_{e-ph} \propto T^{-2}$ from relaxation time measurements of resistance and the rise of the electron temperature [22]. The exact relation for Nb is shown in Fig. 2(A). The resistance variation with temperature for Nb close to $T_c$ is modeled as [23]

$$\rho(T_e) = \frac{\rho_N}{2}\left(1 + \tanh\left(\left(\frac{T_e - T_{sw}}{2a}\right)\left(1 - \tanh\left(\frac{T_e - T_{sw}}{b}\right)\right)\right. \right.$$
$$\left. \left. + \left(\frac{T_e - T_{sw}}{2c}\right)\left(1 - \tanh\left(\frac{T_e - T_{sw}}{b}\right)\right)\right)\right) \quad \mathbf{5}$$

With $a$, $b$ and $c$ values obtained from fits to measured data. $\rho_N$ is the normal state resistivity of the superconducting nanowires, $T_{sw}$ is the current dependent switching temperature and has a functional form of $T_{sw}(J_b) = T_c\sqrt{1 - \left(\frac{J_b}{J_c}\right)^{\frac{2}{3}}}$, The temperature dependence is plotted in Fig. 2(B) showing a close-to-ideal sharp resistance drop with temperature.

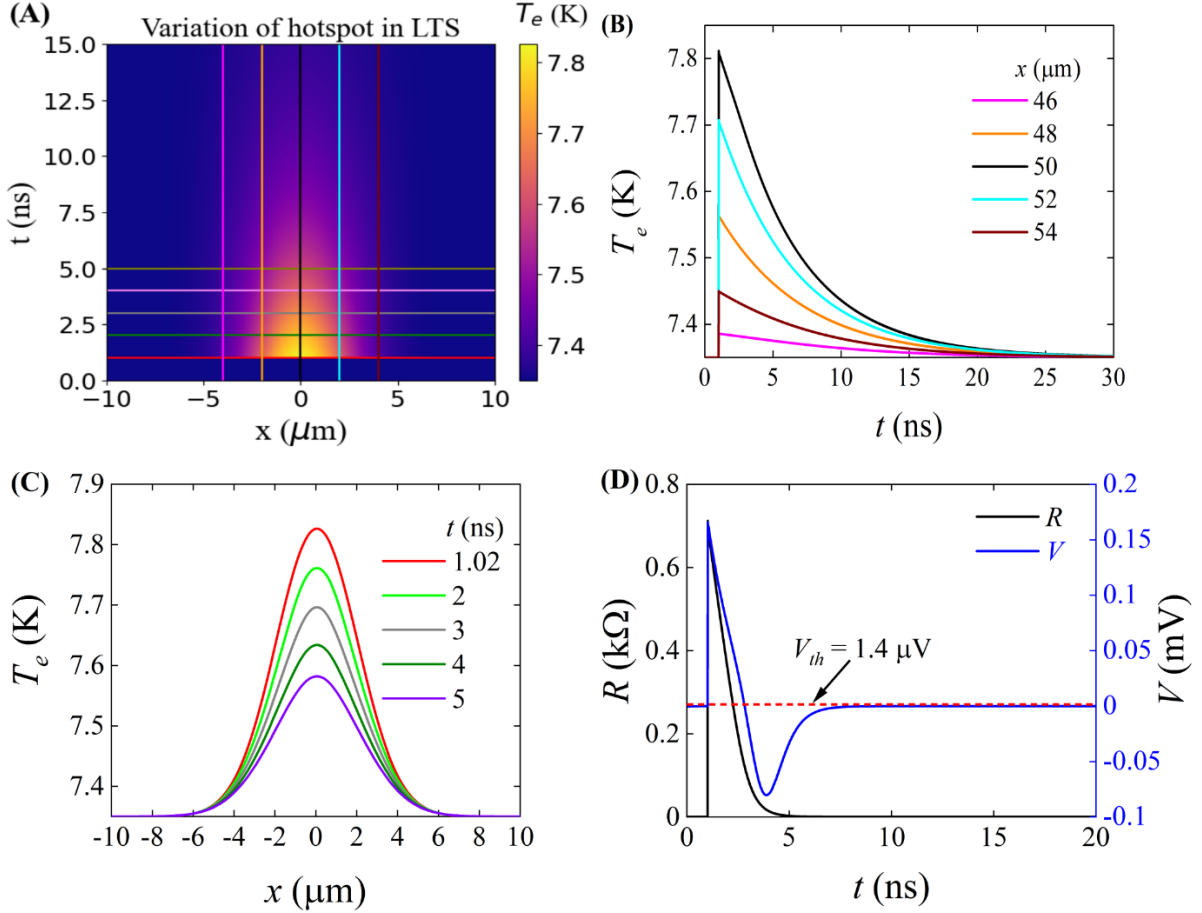

Fig. 3. (A) The spatial and temporal variation of hotspot in LTS, the vertical and horizontal lines represent the temporal and spatial positions for respective variations of $T_e(x)$ and $T_e(t)$; (B) Temporal variation of $T_e$ of LTS at different spaces; (C) Spatial variation of $T_e$ of LTS at different times, we have offset the x-axis so that $x = 0$ be the middle of the photon incidence; (D) Variation of $R(t)$ and $V(t)$ for LTS, the red dashed line represents the thermal noise potential across the 50 Ω shunt resistance.

The biasing conditions is defined by $r$ and $s$ parameters. The wire is taken to be current biased at $J_b = r * J_c$ when the entire sample is set at a temperature $T_b = s * T_{sw}$. A Gaussian photon pulse of width 1 ps with a spatial FWHM of 2 μm is taken to be incident on nanowire such that the focal center of the light source falls at the middle of the wire. The temperature variation with space and time when the nanowire is biased at $r = 0.25$ and $s = 0.95$ is shown in Fig. 3(A) for a representative incident photon pulse strength of 500 photons at 1550 nm. The temporal variation of electron temperature at different spatial positions of the hotspot as marked in Fig. 3(B). The central point of the photon incidence (taken at $x = 0$) reaches a maximum temperature ($T_{e,max}$) of about 7.81 K. While different regions of the wire return to the substrate set temperature at different times, the maximum temporal width of hotspot ($\Delta t$) is found to be 28.28 ns at the photon incidence spot. In Fig. 3(C), we have shown the spatial variation of $T_e(x)$ at different times. The hotspot always remains symmetric around the photon incident point. The maximal spatial width ($\Delta x$) of the hotspot is found to be 14.08 μm at 1.02 ns from the photon incidence. The resistance changes in the nanowire calculated using the equation $\mathbf{5}$ is presented in Fig. 3(D). The hotspot resistance rise time is about 10 ps for a 1 ps photon pulse and decays slowly to the superconducting state in 5 ns. The potential response in the electrical circuit due to the resistance change is calculated using the equation $\mathbf{4}$ and is presented as a function of time in Fig. 3(D). An oscillating potential pulse with a peak value of 2.5 mV is obtained due to the kinetic inductance and parasitic capacitances in the circuitry (nanowire dimensions are presented in Table I). The reset time ($\tau_{reset}$), defined as the time duration after one photon incidence to the detector recovering superconductivity is found to be 7.68 ns for Nb thin nanowire biased at the said conditions. Experimental measurement on a 10 nm thick Nb nanowire biased at $r = 0.5$ and $s = 0.63$ under the irradiation of 404 nm photon source had a reset time of 2 ns. Several other Nb nanowire had been observed to have reset times in the range of $2 - 30$ ns [32]. The closeness to our calculated values provides the confidence that the model with accurate thermodynamic properties can faithfully predict the SNSPD responses.

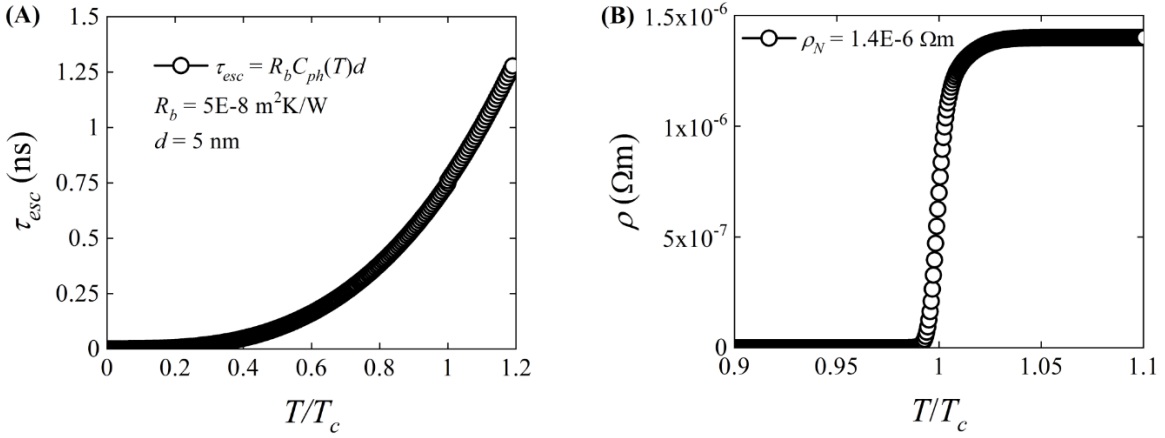

Fig. 4. (A) The variation of heat escape time, $\tau_{esc}(T)$, through phonon to substrate for HTS SNSPD with MgO substrate; (B) Variation of HTS resistivity, $\rho(T)$, as a function of temperature, the fitted parameters are $a = 0.9$ K, $b = 1.08$ K and $c = 1.18$ K.

We now proceed to investigate the responses of a representative HTS for SNSPD applications. YBa$_2$Cu$_3$O$_{7-\delta}$ (YBCO) nanowire is among the popular choices for SNSPD due to the maturity in thin film growth processes and nanostructure fabrications [33]. We have chosen a YBCO nanowire grown on MgO substrate as our model HTS-SNSPD, because MgO substrate is commonly used for YBCO thin film growth. These give us flexibility for future correlation between experimental and theoretical results. The geometry of the nanowire, characteristics of the incident photon are taken to be same for the two materials. The biasing conditions were scaled by the transition conditions. As the pairing mechanism is different in LTS and HTS [34], the different superconducting thermodynamic variables of YBCO are listed in preceding TABLE I. The electronic specific heat and thermal conductivity of HTS are presented in the **supplementary material Section 2**. The HTS thermal propagation characteristics as well as the resistivity transitions vary significantly with the LTS one (compare Fig. 4 with Fig. 2). Notably, the thermal conductivity is found to be smaller in HTS material $\left( \frac{c_{es,Nb}}{c_{es,YBCO}} \Big|_{T=T_c} \sim 23.49 \right)$, and the interface thermal loss factor in the YBCO shows a strong temperature dependence while the Nb is largely invariant. While these aspects are specific details corresponding to Nb and YBCO, it is mentioned to illustrate the capability of the code to handle complex temperature dependences.

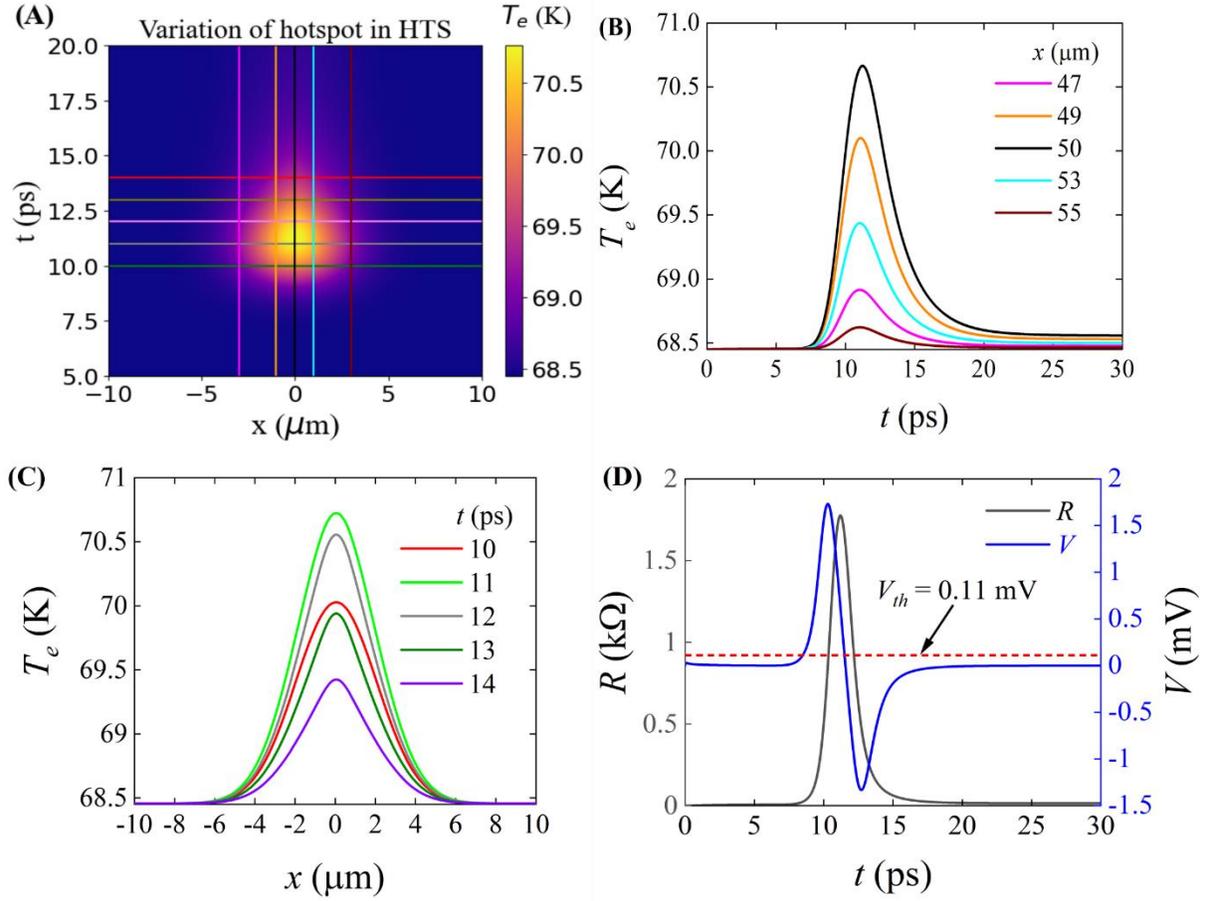

Fig. 5. (A) The spatial and temporal variation of hotspot in HTS, the vertical and horizontal lines represent the temporal and spatial coordinates for respective variations of $T_e(x)$ and $T_e(t)$; (B) Temporal variation of $T_e$ of HTS at different spaces; (C) Spatial variation of $T_e$ of HTS at different times. We have offset the x-axis so that $x = 0$ be the center of photon incidence; (D) Variation of $R(t)$ and $V(t)$ for HTS, the red dashed line represents the thermal noise potential across the 50 Ω shunt resistance.

Similar to the LTS, we consider a specific case where the HTS is biased with $J_b = 0.25J_c$ and the initial temperature is being fixed at $T_b = sT_{sw}$ where $s = 0.96$. The variation of $T_e(x, t)$ of HTS is shown in Fig. 5(A) for the HTS parameters presented in TABLE I at a strength of 1200 photons at 1550 nm. In Fig. 5(B), we have shown the temporal variation of $T_e(t)$ at different spatial points. The maximum temperature, $T_{e,max}$, has been observed the photon incidence point (taken to be $x = 0$). The incidence point takes 11.2 ps to achieve the maximum temperature from the base temperature of 68.45 K. Different special points of the nanowire reaches the base temperature at different time, the middle of the nanowire returns back to the base temperature 14.49 ps after the photon incidence for the same conditions. The maximum temporal width of hotspot ($\Delta t$) is found to be 17.68 ps at $x = 0$ μm. The variation of $T_e(x)$ at different time has been shown in Fig. 5(C). The maximum special width ($\Delta x$) of the formed hotspot in the HTS SNSPD is found to be 13.69 μm at $t = 11$ ps. In Fig. 5(D), we show the change of resistance of the HTS nanowire and the corresponding voltage pulse as a function of time. An exceptionally fast resistance change is observed with the total resistance/potential pulse falling extremely fast to the superconducting state. The reset time $t_{reset}$ is found to be about 10.25 ps. These numbers match closely with experimental reports. The voltage response of the YBCO photosensor (100 nm thick) when irradiated with a femtosecond laser pulse ($\lambda = 790$ nm, 100 fs) is found to recover superconductivity in 1.5 ps [35]. In MgB$_2$ nanowires (35 nm × 120 μm) under a 1560 nm laser irradiation (pulse width = 50 fs, 100 MHz repetition rate) a reset time of about 130 ps is observed [15]. Recently reset time for YBCO microwires (50 nm thick) is reported to be less than 2 ns at 850 nm (4 ps pulse width) [14]. We therefore show for the first time a numerical model that can faithfully reproduce the responses from both LTS and HTS nanowires for arbitrary material properties and incident photon incidence characteristics.

We address the larger question of the fundamental limit on the sensitivity of the nanowires in terms of capability to detect a single photon at different bias points and its dependence on the thermodynamic properties of the material. Towards that, we calculate the sensitivity of the SNSPD in terms of the minimum number of photons required to produce a potential value above the thermal noise floor under different bias conditions. The thermal noise potential over the 50 Ω shunt resistor line is indicated by the dashed red line in the potential plot (Fig. 3(D) and 5(D)). The photon sensitivity as a function of the bias conditions is presented in Fig. 6.

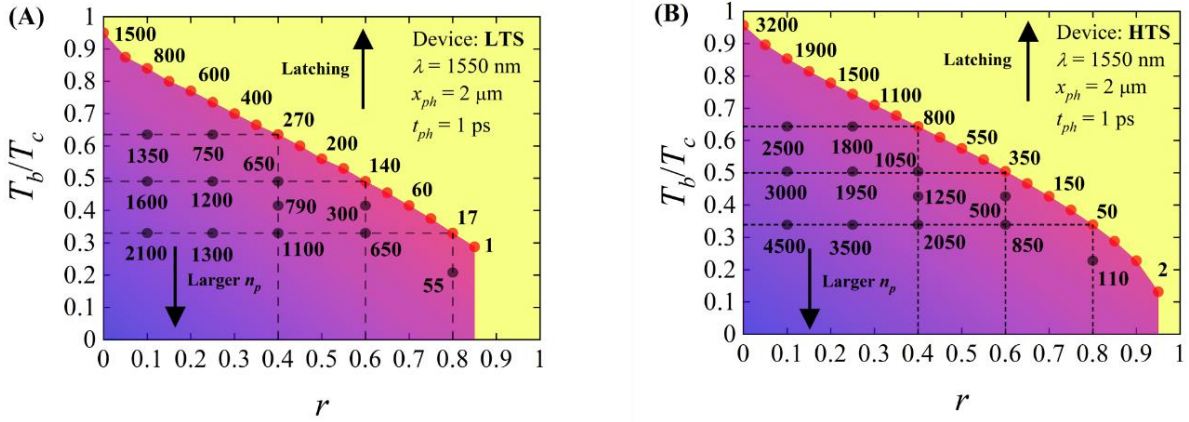

Fig. 6. A phase diagram of $n_p$ as a function of $T_b/T_c$ and $r = J_b/J_c$ for (A) Nb LTS SNSPD and (B) YBCO HTS SNSPD

A bias set point in terms of the current and temperature can lead to two different states: (a) detecting stage where an observable potential pulse is measured for some defined number of photons over the thermal noise floor, (b) latched stage where certain parts of the superconductor get latched in a normal state and never recovers back to photon detection stage. For any sample temperature $T < T_C$, as the biasing current is increased the minimum number of photons ($n_p$) to create a detectable hotspot decreases as noted in the Fig. 6(A). In Fig. 6(A), for LTS-SNSPD we have observed that at a constant $T_b$, with the increase of $r$, the number of photons $n_p$ decreases monotonically (indicated by a dashed line with numbers). The most sensitive current bias point at any temperature can then be obtained for any temperature at the cross-over between the two phases. Similarly at constant $r$, $n_p$ decreases with increase in $T_b$ till the phase interface line is reached. Lastly, when the current bias is increased (increasing $r$), the temperature of cross-over from detection to a latching phase reduces. The slope of the maximum sensitivity line for the LTS is observed to be $d(T_b/Tc)/dr = 0.67$ at $r = 0.25$ and $0.69$ at $r = 0.8$. For a representative Nb thin film taken with a thickness of 5 nm achieves single photon detection capability at 1550 nm for $r = 0.85$, $s = 0.95$ for LTS. The HTS phase diagram also resembles the LTS one. The detection phase and the latching phases are observed with the maximal sensitivity point at its boundary. The maximal sensitivity line has some minor variations. The minimum number of photons required to show a response in a HTS is consistently larger than the LTS case with similar biasing conditions ($r$ and $T_b/T_c$). The slop of the maximal sensitivity line also is not as 'linear-like' as the LTS case and shows drastic changes in the region nearing single photon resolution. A 100 μm YBCO wire is found to achieve a maximum sensitivity an overall maximum sensitivity at a bias condition $r = 0.95$ and $s = 0.96$. There is no recent paper that presents the sensitivity of the photon counting range of a SNSPD, we believe our model will help the designers to design different SNSPD in achieving the maximal sensitivity.

## IV. DISCUSSION

The photon sensitivity is a complex interplay between the thermodynamic variables ($C_{es}, k_{es}$), width of the resistivity transition, the electron-phonon relaxation constants, and several other superconducting parameters. The spatial variation of hotspots is found to be 14.08 μm and 13.69 μm for $r = 0.25$ and $s = 0.95$ for LTS and HTS respectively. Although the special variation of hotspot remained almost same, whereas the temporal variation ($\Delta t$) of the hotspot for LTS and HTS were found to be 28.28 ns and 17.68 ps respectively. The higher $k_{es}$ of the LTS and poorer $\tau_{esc}$ allows for trapping the photo generated heat inside the superconductor. and driving hotspot growth, on contrary relatively lower $k_{es}$ and faster $\tau_{esc}$ and $\tau_{e-ph}$ of HTS promotes faster growth of hotspot and dissipate the excess heat into substrate which helps the device to return to its base condition, which results the HTS SNSPD to show smaller $\Delta t$ than LTS SNSPD.

The spatial variation of hotspot ($\Delta x$) is a complex interplay between $C_{es}, k_{es}, \Delta x, \Delta T, \Delta H$ and $\Delta t$; where $\Delta T = T_{e,max} - T_b$ and $\Delta H$ is the heat required for temperature raise, $\Delta T$. The relationship is known to follow, $k_{es} A \Delta T / \Delta x = \Delta H / \Delta t$. The temperature differences ($\Delta T$) for LTS and HTS SNSPD for the aforesaid conditions are found to be 0.46 K and 2.19 K respectively. As $C_{es,LTS} > C_{es,HTS}$, the heat required the temperature is higher for LTS than HTS, but the amount of heat transferred ($\Delta H / \Delta t$) is higher for HTS SNSPD. Therefore, the competitions between the thermodynamic variables make the approximate hotspot size $\Delta x$ to be almost same for both the materials.

Lastly, we address sensitivity and the response times. To illustrate the information, we calculated the maximum sensitivity plots for an additional excitation wavelength ($\lambda = 830$ nm) for comparison.

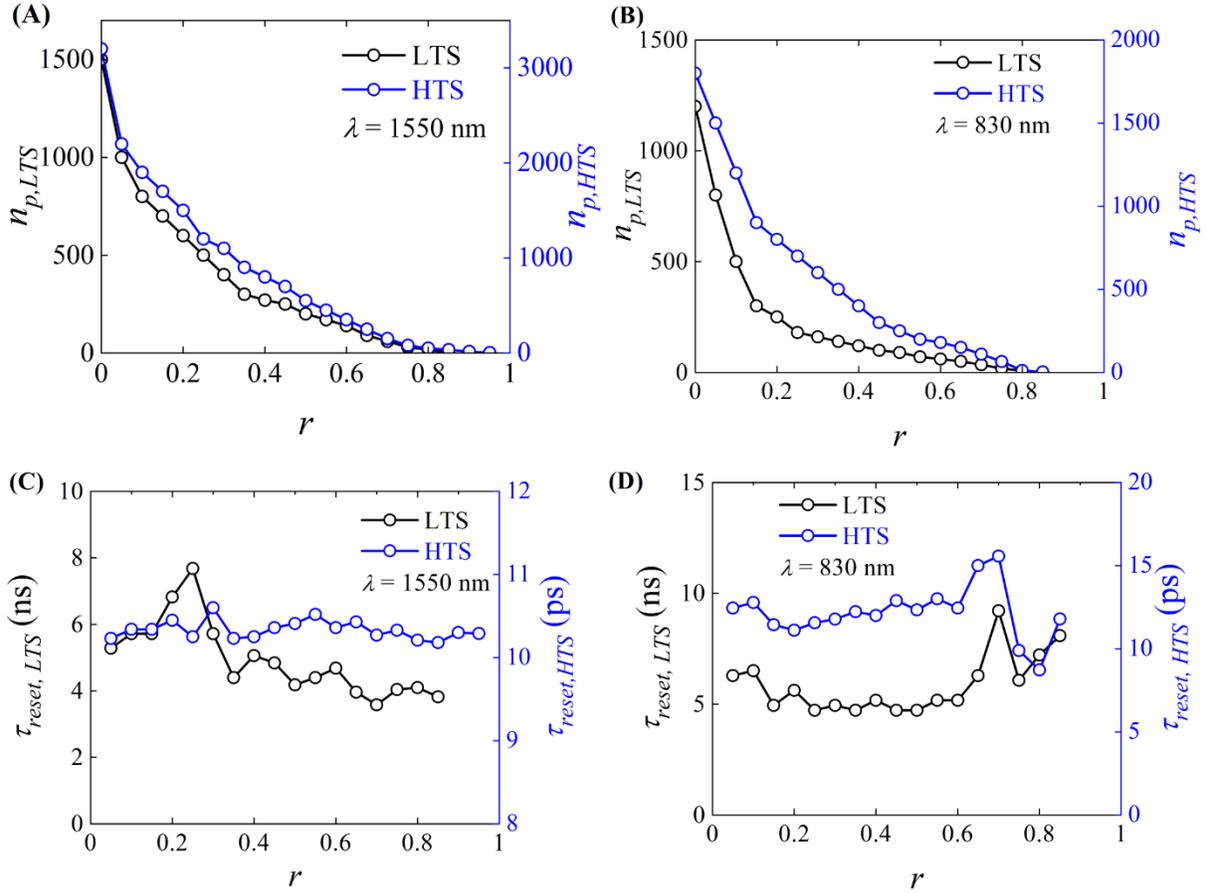

Fig. 7. Number of photons $(n_p)$ as a function of the ratio between biasing and critical current densities $(r = J_b/J_c)$ for LTS ($s = 0.95$) and HTS ($s = 0.96$) based SNSPD at (A) $\lambda = 1550$ nm and (B) $\lambda = 830$ nm. $\tau_{reset}$ as a function of the ratio between biasing and critical current densities $(r = J_b/J_c)$ for LTS ($s = 0.95$) and HTS ($s = 0.96$) based SNSPD at (C) $\lambda = 1550$ nm and (D) $\lambda = 830$ nm.

The $n_p(r)$ variation at bias temperature of ($s = 0.95$) for LTS and HTS is presented in Fig. 7(A) and 7(B) for two different wavelengths respectively. Under all conditions, the LTS always had a better photon number sensitivity than the HTS. The sharper resistivity transition of the LTS (HTS transition width is four times the LTS width) is largely responsible for the higher sensitivity. As the biasing current density increases, $n_p$ decreases for both LTS and HTS monotonically. Even though, we observed the ultimate photon sensitivity for both the SNSPDs at higher biasing current($J_b \geq 0.8J_c$), the variations of $n_p(r)$ was found to almost similar between LTS and HTS for the 1550 nm while there is noticeable difference for the smaller wavelength photon. The number of photons needed for destroying localized superconductivity is roughly proportional to $hc/\lambda\Delta(T)$. The photon associated with 830 nm wavelength has almost two times higher energy than the 1550 nm-photon, whereas the order parameter of Nb is about an order of magnitude smaller than YBCO. Hence, the lower wavelength photon with a higher energy generates significantly higher number of excited carriers in the smaller order parameter Nb than in the YBCO thin film making it more noticeably sensitive. On the other hand, the reset times as seen in Fig. 7(C) and 7(D) show no much variation either with the bias condition nor with the wavelength of the incident photon. However, the important observation is that the HTS SNSPD has three order of magnitude faster response than its LTS counterpart. The SNSPD temporal response depends strongly on the $\Delta t$, $\tau_{e-ph}$, and the kinetic inductance in addition to the thermodynamic variables mentioned above [18]. The HTS has a much smaller kinetic inductance value due to the smaller magnetic penetration depths (Nb and YBCO are found to be 39 nm and 26 nm respectively [36,37]). The calculated kinetic inductance (from equation 4) for LTS and HTS are 0.19 pH and 0.085 pH respectively. The smaller inductance of the HTS in conjunction with the much faster thermal escape rates into the substrate plays a crucial rule in the temporal response of HTS SNSPDs. Under the excitation of different wavelengths, the HTS-SNSPD is always found to be faster than LTS-SNSPD. Therefore, the aforesaid observations support that our modified model can handle a range of diverse experimental scenarios and we believe that it can be used as a primary tool to quantify performance of basic different SNSPDs.

## V. CONCLUSION

A generalized two temperature model developed here calculates the the hotspot dynamics of different SNSPDs. We have calculated the hot-spot dynamics of both LTS and HTS SNSPD numerically and find the results to match closely with the experimental results. We show a phase-diagram like plot to estimate the minimum photon count required to observe a signal above the noise limit in Nb and YBCO as an example. Our calculations show that both LTS and HTS can resolve single photons. Even though the HTS nanowire is three orders of magnitude faster than the LTS nanowire, the LTS is more sensitive for the same nanowire dimension and bias conditions. We believe that with appropriate expressions for the various physical properties, the sensitivity and response of different superconducting nanowires can be calculated apriori using the numerical treatment mentioned here.

## Acknowledgment

The authors also acknowledge the members of the Neos Lab, Department of Material Science and Engineering, Indian Institute of Technology Delhi, New Delhi 110016, India for meaningful academic discussions.